\documentclass[aip,graphicx]{revtex4-1}
\usepackage{alltt,color}
\usepackage{graphicx}
\usepackage{epstopdf}
\usepackage{float}
\usepackage{amsmath}
\usepackage{cases}
\usepackage{multirow}
\usepackage{appendix}
\usepackage{array}
\usepackage{bm}

\makeatletter

\newcommand{\Rmnum}[1]{\expandafter\@slowromancap\romannumeral #1@}
\makeatother

\newcommand{\aap}{    {\it Astron. Astrophys.}}
\newcommand{\grl}{    {\it Geophys. Res. Lett.}}
\newcommand{\planss}{   {\it Planetary Space Science}}

\newcommand{\ssr}{    {\it Space Sci. Rev.}}

\draft 

\begin{document}


\title{Electron acceleration by turbulent plasmoid reconnection} 



\author{X. Zhou}
\email[]{zhou@mps.mpg.de}
\altaffiliation[Also at ]{Key Laboratory of Dark Matter and Space Astronomy,
Purple Mountain Observatory, Chinese Academy of Sciences, Nanjing,
210008, China.}
\author{J. B\"{u}chner}
\author{F. Widmer} 
\altaffiliation[Presently at ]{CEA, IRFM, F-13108 Saint Paul-Lez-Durance, France}
\author{P. A. Mu\~{n}oz}
\affiliation{Max Planck Institute for Solar System Research,
Justus-von-Liebig-Weg 3, 37077 G\"{o}ttingen, Germany}
%


\date{\today}

\begin{abstract}
In space and astrophysical plasmas, like in planetary magnetospheres, as
that of Mercury,energetic electrons are often found near current sheets (CSs),
which hints at electron acceleration by magnetic reconnection. Unfortunately,
electron acceleration by reconnection is not well understood, yet.
In particular, acceleration by  turbulent plasmoid reconnection.
We have investigated electron acceleration by turbulent plasmoid reconnection,
described by MHD simulations, via test particle calculations.
In order to avoid resolving all relevant turbulence scales down to the
dissipation scales, a mean-field turbulence model is used to describe the
turbulence of sub-grid scales (SGS) and their effects via a turbulent
electromotive force (EMF). The mean-field model describes the turbulent
EMF as a function of the mean values of current density, vorticity, magnetic
field as well as of the energy, cross-helicity and residual helicity of
the turbulence. We found that, mainly around X-points of turbulent
reconnection, strongly enhanced localized EMFs most efficiently accelerated
electrons and caused the formation of power-law spectra.
Magnetic-field-aligned EMFs, caused by the turbulence, dominate the
electron acceleration process. Scaling the acceleration processes to
parameters of the Hermean magnetotail, electron energies up to 60 keV
can be reached by turbulent plasmoid reconnection through the thermal plasma.
\end{abstract}


\maketitle 

\section{Introduction}
\label{Introduction}

Energetic electrons are observed remotely and in-situ
throughout the whole solar system and beyond. They are accelerated during
solar flares as well as in planetary magnetospheres and so on.
Since high-energy electrons are often found near current sheets (CSs),
magnetic reconnection is thought to be one key process for their acceleration.
On the other hand, collisionless space plasmas are usually highly turbulent.
The consequences of turbulence and magnetic reconnection for
electron acceleration, however, are not well understood, yet.

Different models have been proposed to take into account
turbulence in magnetic reconnection, e.g.,
Refs.~\onlinecite{Matthaeus&Lamkin1985PhFl, Lazarian&Vishniac1999ApJ} and
test particle studies have been carried out to investigate the particle
acceleration resulting from MHD-turbulent magnetic reconnection.
\emph{Ambrosiano et al.} \cite{Ambrosiano1988} found efficient
particle energization by two dimensional (2D) turbulent magnetic reconnection.
\emph{Dmitruk et al.} \cite{Dmitruk2004} carried out test particle studies in
3D turbulent magnetic reconnection, which revealed a
preferential electron acceleration parallel to the magnetic field in
localized current sheets.
\emph{Petkaki $\&$ MacKinnon} \cite{Petkaki2007,Petkaki2011},
\emph{Burge et al.} \cite{Burge2012} analysed the consequences of turbulent
electromagnetic fields on particle acceleration at X-type neutral points.
They found an increasing energization in strong turbulence and formation
of bi-modal (double-peak) distributions.
\emph{Kowal et al.} \cite{Kowaletal2012PhRvL} studied proton acceleration in
3D turbulent CSs.
They found that the proton acceleration rate was highly enhanced by a
first-order Fermi process due to contracting magnetic fluctuations.
These prior calculations, however, did not consider the effects of
turbulence in sub-grid-scales due to the limitation of the computing resources.
\emph{Yoshizawa} \cite{Yoshizawa1990PhFlB} proposed a mean-field
turbulence model to describe the effects of small-scale turbulence on
the large-scale dynamics. Recently, \emph{Yokoi $\&$ Hoshino} \cite{Yokoi&Hoshino2011PhPl}
suggested their consequences to turbulent magnetic reconnection via an electromotive force (EMF).
The EMFs in this mean-field turbulence model are related to the
mean current density, vorticity, magnetic field
as well as to the energy, cross-helicity and residual helicity of the MHD
turbulence.
In the present work, we use this mean-turbulence model to investigate
electron acceleration by turbulent plasmoid reconnection with parameters
of the Hermean magnetotail. Energetic electrons have been observed
in-situ in the turbulent plasma of the Hermean magnetotail, their
acceleration mechanisms are, however, less understood compared to the
energetic electrons in the Earth's magnetotail \cite{Birnetal2012SSRv}.

Mercury is the closest planet to the Sun with a quite weak and small magnetosphere.
On the other hand, it has the most dynamical magnetosphere among all the four
terrestrial planets.
Several spacecrafts have been devoted to investigate the environment of
Mercury: Mariner 10, the two Helios spacecrafts, the MErcury Surface,
Space ENvironment, GEochemistry and Ranging (MESSENGER) mission with its
Energetic Particle Spectrometer (EPS), the X-Ray Spectrometer
(XRS), the Gamma-Ray and Neutron Spectrometer (GRNS) as well as the Gamma Ray
Spectrometer (GRS).
The first in-situ measurements of the energetic particles at Mercury were made by
Mariner 10 in 1974-75. Mariner 10 discovered that energetic electron bursts
in the Hermean magnetotail have a time duration of about 10 s \cite{Eraker&Simpson1986JGR}.
Pileup in the instrument electronics of the Mariner 10, however, led
overestimations of the particle energies \cite{Armstrongetal1975JGR}.
Later, the MESSENGER (starting 2011) regularly observed energetic electrons with
energies up to 100-200 keV in the Hermean magnetotail,
while the typical particle energy in the upstream solar wind are
typically only 1.5 - 10 keV based on the observation of the Helios spacecraft
\citealp{Kirsch&Richter1985AnGeo...3...13K}.
Observations of the MESSENGER provide a strong evidence for electron
acceleration in the Hermean magnetotail.
The MESSENGER, however, did not find energetic ions
\citealp{Ho2011a, Hoetal2012JGRA, Lawrenceetal2015JGRA, Bakeretal2016JGRA}.
Various acceleration mechanisms have been proposed to explain these observed
high electron energies in the Hermean magnetotail: inductive
acceleration via substorm-like dipolarization \citealp{Delcourtetal2005},
stochastic acceleration, wave-particle interactions, bow-shock
energization, and magnetic reconnection in the magnetotail
(see \emph{Zelenyi et al.} \citealp{Zelenyietal1990JGR, Zelenyietal2007}).
%
MESSENGER discovered not only clear signatures of energetic electrons
but also multiple plasmoids and plasmoid reconnection in the Hermean
magnetotail \citealp{Slavin2009, DiBraccioetal2015P&SS, Sunetal2016JGRA}.
We followed these observations using characteristic parameters of the
Hermean magnetotail, plasmoid reconnection and plasma turbulence, since
plasmas with high Reynolds number, typical for astrophysical environments,
are prone to be turbulent \citealp{Matthaeus&Lamkin1985PhFl}.

Theoretical analyses \citealp{Loureiroetal2007, Uzdenskyetal2010},
observations \citealp{Hoshinoetal1994, Karlicky2004},
MHD and particle-in-cell (PIC)-code simulations \citealp{Bartaetal2011,Karlickyetal2012}
have shown that fast magnetic reconnection can be due to plasmoid instabilities
forming magnetic islands (or flux ropes) in elongated current sheets with
finite guide field in the direction perpendicular to the reconnection plane.
To describe the influence of turbulence in sub-grid-scales on plasmoid reconnectoin,
\emph{Widmer et al.} \cite{Widmeretal2016PhPl} recently used the
the mean-field turbulence model of
\emph{Yoshizawa} \cite{Yoshizawa1990PhFlB} and
\emph{Yokoi $\&$ Hoshino} \cite{Yokoi&Hoshino2011PhPl}.
They found that turbulence can enhance
reconnection rates in dependence on the properties of the turbulence.

Electron acceleration by plasmoid reconnection has been investigated previously via
theoretical analyses \citealp{Drakeetal2006Natur, Drakeetal2013ApJ}
and PIC code simulations \citealp{Drakeetal2006Natur, Okaetal2010ApJ, Tanakaetal2011PhPl}.
Those studies revealed the importance of the curvature and gradient drifts by
contraction and coalescence of plasmoids.
Test particle calculations \citealp{Zhouxiaowei2015ApJ, Guidonietal2016ApJ, Borovikovetal2017ApJ}
partially confirmed this prediction, however, without
considering the consequences of turbulence.
In this study, we investigate the consequences of EMF caused by turbulence on
electron acceleration in plasmoid-unstable CSs by using the mean-field turblence model of
\emph{Yoshizawa} \cite{Yoshizawa1990PhFlB} and
\emph{Yokoi $\&$ Hoshino} \cite{Yokoi&Hoshino2011PhPl}.
The relevant mean electromagnetic fields are obtained by a Gaussian spatial
filtering \cite{Schmidt2015LRCA, Widmeretal2016PhPl}.
We carried out test particle calculations in the macroscopic (mean)
fields based on a relativistic guiding center approximation.

This paper is organized as follows.
In Section \ref{MHD-Model}, we describe the high resolution MHD
simulations of plasmoid-unstable CSs and the mean-field turbulence model of
\emph{Yoshizawa} \cite{Yoshizawa1990PhFlB} and
\emph{Yokoi $\&$ Hoshino} \cite{Yokoi&Hoshino2011PhPl}.
In Section \ref{Results}, we reveal the resulting electron acceleration in
plasmoid-unstable CSs , discriminate the different physical effects
on electron energization and localize the electron acceleration sites.
Characteristic trajectories of the
strongest energized electrons as well as the resulting electron energy spectra
are provided in Section \ref{Results}.
Conclusions and discussions are contained in Section \ref{Conclusions}.

\section{Plasmoid Reconnection and Turbulence}
\label{MHD-Model}


 \subsection{Plasmoid reconnection}
 \label{MHD Simulation}

In this section we present the MHD simulations
that we used for our electron acceleration
calculations (see also of \emph{Widmer et al.} \cite{Widmeretal2016PhPl}). Our MHD simulations
describe the evolution of the electromagnetic fields of
plasmoid-unstable CSs by solving the following set of resistive MHD equations:
\begin{align}
\displaystyle\frac{\partial \rho}{\partial t} & =  - \nabla \cdot (\rho \vec{U})
\label{MHDEquations1} \\
\displaystyle\frac{\partial \rho \vec{U}}{\partial t} &= - \nabla \cdot [  \rho \vec{U} \vec{U}
                                                                       + \displaystyle\frac{1}{2}(p+\textcolor{black}{\bm{\displaystyle\frac{B^2}{\mu_{0}}}})\text{I}
                                                                       - \textcolor{black}{\bm{\displaystyle\frac{\vec{B}  \vec{B}}{\mu_{0}}}}  ] 
                                                      + \chi \nabla^{2} \rho \vec{U}
\label{MHDEquations2} \\
\displaystyle\frac{\partial \vec{B}}{\partial t}   &=  \nabla \times \left( \vec{U} \times \vec{B} -  \eta \vec{J} \right)
\label{MHDEquations3} \\
\displaystyle\frac{\partial h}{\partial t} &= - \nabla \cdot (h \vec{U})
                                              + \displaystyle\frac{\gamma_{0}-1}{\textcolor{black}{\bm{2}} \gamma_{0}h^{\gamma_{0}-1}} \eta J^{2} +  \chi \nabla^{2} h,
\label{MHDEquations4}
\end{align}
using the GOEMHD3 code \citealp{Skalaetal2015A&A, Widmeretal2016PhPl}.
Here $\rho$ is the mass density, $\vec{U}$ is the plasma velocity,
$\text{I}$ is the three-dimensional identity matrix, $\vec{B}$ is the magnetic field
and $h$ is related to the thermal pressure $p$ via $p=2h^{\gamma_{0}}$,
with $\gamma_{0}=5/3$ being the ratio of specific heats in adiabatic conditions.
Amp\`{e}re's law is used to compute the current density
$\vec{J}=\nabla \times \vec{B}/\mu_{0}$, with $\mu_{0}$ being the
vacuum magnetic permeability.
A small homogeneous (normalized) resistivity $\eta=0.001$ corresponds to a
sufficiently large magnetic Reynolds number, so that a plasmoid instability
takes place \citealp{Widmer2015PhPl}
(see below for the normalization of the resistivity).
$\chi$ in Eqs.~\eqref{MHDEquations2} and \eqref{MHDEquations4} represents
the amount of local viscosity switched on as soon as any
numerical instability starts.
\textcolor{black}{\text{The variables in Eqs.~\eqref{MHDEquations1} to \eqref{MHDEquations4} are
dimensionless.}}
In order to be applied to the typical Mercury conditions,
\textcolor{black}{\text{The variables in Eqs.~\eqref{MHDEquations1} to \eqref{MHDEquations4}}}
are normalized to values characteristic for the Hermean magnetotail
\citealp{Zelenyietal1990JGR, Zelenyietal2007, Fujimotoetal2007SSRv} '
\textcolor{black}{\text{in the MHD simulation:}}
the length scale is the assumed halfwidth of the Hermean magnetotail CS
$L_{0} = 2.5 \times 10^{4}$ $m$,
the magnetic field is $B_{0} = 7.5 \times 10^{-8}$ $T$,
and the typical Alfv\'{e}n transit time is $t_{0}=1$ $s$.
All other normalizations can be derived from the above three normalizing quantities:
velocities are normalized to $V_{0}=L_{0}/ t_{0} = 2.5 \times 10^{4}$ $m/s$,
electric fields to $E_{0}=V_{0} B_{0} = 1.9 \times 10^{-3}$ $V/m$,
the current densities to $J_{0}=B_{0}/(\mu_{0}L_{0}) = 2.4 \times 10^{-6}$  $A/m^{2}$
and the resistivity to $\eta_{0}=\mu_{0} L_{0} V_{0} = 7.9 \times 10^{2}$ $\Omega \cdot m$.

In these simulations, an almost two-dimensional simulation
box containing $4 \times 3200 \times 12800$ grid points in a domain
$L_X \times L_Y \times L_Z = 0.4 \times 80 \times 320\, L_{0}^{3}$, where
$X, Y$ and $Z$ are the directions out of the CS plane, across and along the CS,
respectively.
The MHD simulation uses periodic boundary conditions in the $Y$ and $Z$ directions.
Double Harris-type CSs with a finite guide field $b_g=B_{x0}/B_{0}$ are
initialized as follows (in dimensionless units):
\begin{align}
      \vec{B} &=b_{g} \overrightarrow{e_{x}} + \left[ \tanh(y+d)-\tanh(y-d)-1 \right] \overrightarrow{e_{z}}
\label{Ini_1} \\
      h &=\displaystyle\frac{1}{2} \left(1+\beta_{p}-B^{2} \right)^{1/\gamma_{0}}
\label{Ini_2}
\end{align}
Here we will further present results obtained for a guide magnetic field $b_g = 2$.
 $\beta_{p}=0.5$ is
the plasma-$\beta$ (ratio between thermal
and magnetic pressures), the half-distance between the two CSs is $d=20$ and the
symbols $\overrightarrow{e_{j}}$ denote the unit vectors in the directions
$j=x, y$ or $z$.
A multi-mode initial perturbation spectrum is used to trigger the plasmoid
instability:
\begin{align}
      B_{y}(z) &= \sum^{128}_{k=1}0.01 \lambda_{1} \sin\left[2 \pi k \left(\displaystyle\frac{z}{L_{z}}+ \lambda_{2}\right)\right]
\label{Perturbation}
\end{align}
where $\lambda_{1}$ and $\lambda_{2}$ are random numbers in the range $[0,1]$ \cite{Widmeretal2016PhPl}.

 \subsection{Mean-field approach of turbulence}
 \label{Mean-Field Approach}

In a mean-field approach \cite{Krause1980, Schmidt2015LRCA}, the induction
equation (Eq.~\ref{MHDEquations3}) becomes
\begin{align}
      \partial_t\overline{\vec{B}}= - \nabla \times \overline{\vec{E}}
                                  = \nabla \times \left( \overline{\vec{U}} \times \overline{\vec{B}} - \eta \overline{\vec{J}} - \overrightarrow{\varepsilon_{M}} \right)
\label{Equ0}
\end{align}
where $\overrightarrow{\varepsilon_{M}}$ denotes the EMF, due to the turbulence.
We apply the turbulence model, proposed by
\emph{Yoshizawa} \cite{Yoshizawa1990PhFlB} and
\emph{Yokoi $\&$ Hoshino} \cite{Yokoi&Hoshino2011PhPl} to obtain
$\overrightarrow{\varepsilon_{M}}$.
In that model, the EMF $\overrightarrow{\varepsilon_{M}}$
is obtained as a function of the
mean current density $\overline{\vec{J}}$,
the vorticity $\overline{\vec{\Omega}}$ ($\vec{\Omega} = \nabla \times \vec{U}$) and
the mean magnetic field $\overline{\vec{B}}$:
\begin{align}
       \overrightarrow{\varepsilon_{M}} =
                                           \textcolor{black}{\bm{\mu_{0}}} \beta \overline{\vec{J}}
                                           - \omega \textcolor{black}{\bm{\sqrt{\mu_{0} \overline{\rho}}}} \overline{\vec{\Omega}}
                                          - \alpha \overline{\vec{B}}
\label{Equ2}
\end{align}
In the model of \emph{Yoshizawa} \cite{Yoshizawa1990PhFlB} and
\emph{Yokoi $\&$ Hoshino} \cite{Yokoi&Hoshino2011PhPl}, the $\beta$ term in the EMF plays the role of
a turbulent diffusion while the $\alpha$ term corresponds to the usual dynamo
term that accounts for a possible magnetic field generation or at least
its sustainment against diffusion. It appears due to the inhomogeneous flow effects on turbulence.
Coupled with the vorticity $\vec{\Omega}$, the $\omega$ term accounts for the contribution
of the the vorticity to the EMF. Depending on its relative signature compared to the $\beta$ term,
this term might act as an additional dynamo term \cite{Yokoietal2016ApJ...824...67Y}.
A complete description of Eq.~\eqref{Equ2} as well as the physical meaning of the
coefficients \textcolor{black}{\text{$\beta, \omega$ and $\alpha$}}
 can be found in Ref.~\onlinecite{Yokoi2013GApFD.107..114Y}.
According to \emph{Yoshizawa} \cite{Yoshizawa1990PhFlB} and
\emph{Yokoi $\&$ Hoshino} \cite{Yokoi&Hoshino2011PhPl}, the
coefficients $\beta, \omega$ and $\alpha$ in Eq.~\eqref{Equ2} are
related to the  (normalized)
turbulent energy density $K$, cross-helicity density $W$ and
residual helicity $H$ as
\begin{align}
      \beta &= \tau C_{\beta} K,   \ \ \ \  \omega = \tau C_{\omega} W,   \ \ \ \  \alpha= \tau C_{\alpha} H
\label{Equ2_1}
\end{align}
where $C_{\beta}, C_{\omega}$ and $C_{\alpha}$ are constants of the order of
$10^{-2} - 10^{-1}$ $\citealp{Hamba1990PhFlB...2.3064H, Yoshizawa1990PhFlB, Hamba1992PhFl, Higashimorietal2013PhRvL}$ and
$\tau$ is the characteristic decay time of the turbulence.
In a lowest order approximation, $\tau$ can be considered
to be constant and of the order of the initial Alfv\'{e}n transit time
($\tau=1$ or $2$ $t_{0}$), i.e., the time needed for
an Alfv\'{e}n wave to cross the initial CS.
This is a simplified approach for $\tau$ based on previous studies
of magnetic reconnection which revealed the highest reconnection rates.
In their configurations, the fastest magnetic reconnection due to mean-field
turbulence effects was found for $\tau \sim 1.0 - 2.0$
(e.g., Refs.~\onlinecite{Higashimorietal2013PhRvL, Widmer2015PhPl}).
Turbulent energy $K$, cross-helicity $W$ and residual helicity
$H$ are defined as
\begin{align}
      K &= \displaystyle\frac{1}{2}\left[ \left(\overline{U^{2}}-\overline{U}^{2} \right)
                                         +\displaystyle\frac{\overline{B^{2}}-\overline{B}^{2}}{\textcolor{black}{\bm{\mu_{0}}}\overline{\rho}}\right]
      \nonumber \\
      W &= \displaystyle\frac{\overline{\vec{U} \cdot \vec{B}}-\overline{\vec{U}} \cdot \overline{\vec{B}}}{\textcolor{black}{\bm{\sqrt{\mu_{0} \overline{\rho}}}}}
      \nonumber \\
      H &=- \left( \overline{\vec{U} \cdot \vec{\Omega}}-\overline{\vec{U}} \cdot \overline{\vec{\Omega}} \right)
          + \displaystyle\frac{\overline{\vec{B} \cdot \vec{J}}-\overline{\vec{B}} \cdot \overline{\vec{J}}}{\overline{\rho}}
\label{Equ3}
\end{align}
Combining Eqs.~\eqref{Equ0} to \eqref{Equ2},
the mean electric field $\overline{\vec{E}}$ reads
\begin{align}
      \overline{\vec{E}} &= - \overline{\vec{U}} \times \overline{\vec{B}} + \eta \overline{\vec{J}} + \overrightarrow{\varepsilon_{M}}    \nonumber\\
                         &= - \overline{\vec{U}} \times \overline{\vec{B}}
                               + (\eta+\textcolor{black}{\bm{\mu_{0}}}\beta)\overline{\vec{J}}
                               - \omega \textcolor{black}{\bm{\sqrt{\mu_{0} \overline{\rho}}}} \overline{\vec{\Omega}}
                               - \alpha \overline{\vec{B}}
\label{ElectricField}
\end{align}
In this mean-field approach, for plasmas with high magnetic Reynolds numbers
(small $\eta$), annihilation of the magnetic fluxes is solely due to the turbulence
by the $\beta$-related term, which allows the collisionless plasmas to have
a possibility to undergo magnetic reconnection.
In addition, the $\omega-$ and $\alpha-$terms lead to (dynamo-)generation (or
sustainment) of magnetic fields \citealp{Yokoi&Hoshino2011PhPl}.

We use a filter function $\Gamma(\vec{r}, \vec{r}')$ to obtain the mean
component $\overline{F}$ of a quantity $F$:
\begin{align}
      \overline{F(\vec{r})} = \int \Gamma(\vec{r},\vec{r}') F(\vec{r}') d \vec{r}'
\label{FFF1}
\end{align}
For the filtering, we choose a Gaussian filter with a kernel function
\citealp{Schmidt2015LRCA, Widmeretal2016PhPl}:
\begin{align}
      \Gamma(\vec{r},\vec{r}')=\left( \displaystyle\frac{6}{\pi \Delta^2} \right)^{3/2}
                                 \exp \left[ - \displaystyle\frac{(\vec{r} - \vec{r}')^2}{\Delta^2 /6}\right]
\label{FFF2}
\end{align}
where $\Delta$ is the filter width. For $\Delta=4$ (grid size), the Reynolds rules
\citealp{Krause&Raedler1980opp} ($\overline{\overline{F}}=\overline{F}$, $\overline{f'}=0$ and
$\overline{f'\overline{F}}=0$, here $F=\overline{F}+ f'$) are satisfied best.
From here onwards, the variables used for the electromagnetic fields
will be replaced by their mean values and the symbol '$\overline{\ast}$'
for the mean fields will be omitted.

According to the mean-field turbulence theory (see Eq.~\ref{ElectricField}),
$C_{\beta}>C_{\omega}$ and $C_{\beta}>C_{\alpha}$ cause diffusion
rather than generation of magnetic fields.
In this case, the turbulent EMF $\overrightarrow{\varepsilon_{M}}$ will
accelerate particles.
Larger $C_{\beta}$ and $\tau$ correspond to a larger
EMF $\overrightarrow{\varepsilon_{M}}$, which causes fast magnetic flux
annihilation and reconnection \cite{Yokoi&Hoshino2011PhPl, Widmeretal2016PhPl}.
We use $C_{\omega}=0.04$, $C_{\alpha}=0.001$ (see \emph{Widmer et al.}
\cite{Widmeretal2016PhPl}) and investigate the influence of the turbulence on
electron acceleration by varying $C_{\beta}$ (=0.05, 0.5, 1.0) and $\tau$ (=1.0, 2.0), see
Table~\ref{tab1},
due to the
$\beta \propto \tau C_{\beta} $ term in Eq.~\eqref{Equ2} plays the role of transporting
energies from the magnetic fields to the plasma and particles.
We varied the turbulence model parameters because the actual valid
theoretical values for $C_{\beta}$ and $\tau$ have not been derived, yet. What is known from
the turbulence theory is that $C_{\beta}$ should be of the order of $O(10^{-2} - 10^{-1})$.
For $\tau$, it is known that more efficient reconnection is for
$\tau \sim 1.0 - 2.0$~\cite{Higashimorietal2013PhRvL, Widmer2015PhPl}.
Case A with $C_{\beta}, \tau = 0$ is chosen for comparison with the limited case where turbulence
is neglected, i.e., laminar reconnection~\citealp{Watsonetal2007PhPl} where
the magnetic flux annihilation is not affected by turbulence.
\begin{table}[tp]
\begin{center}
\begin{tabular}{|p{2cm}<{\centering}|p{2cm}<{\centering}|p{2cm}<{\centering}|p{2cm}<{\centering}|p{2cm}<{\centering}|p{2cm}<{\centering}|}
  \hline
  Case        & A &   B  &  C  &  D  &  E  \\
  \hline
  $C_{\beta}$ & 0 & 0.05 & 0.5 & 0.5 & 1.0 \\
  \hline
  $\tau$      & 0 & 1.0  & 1.0 & 2.0 & 1.0 \\
  \hline
\end{tabular}
\end{center}
\caption{Turbulence parameters used in our calculations.}
\label{tab1}
\end{table}
Note that $C_{\beta}$ in Case E is out of the range $10^{-2} - 10^{-1}$. This
investigation allows to reveal the importance of the $\omega-$
and $\alpha-$terms (Eq.~\eqref{ElectricField}) comparing its outcome with
that of Cases D and E. The coefficients $\tau C_{\alpha}$ and $\tau C_{\omega}$
in Case D are two times as large as those of Case E for the $\omega-$ and
$\alpha-$terms, while their coefficient $\tau C_{\beta} = 1$ for the $\beta-$term
are same.
Note that the $\beta-$term can be balanced by the $\omega-$ and $\alpha-$terms
\citealp{Yokoi&Hoshino2011PhPl, Widmeretal2016PhPl}.
Larger $\omega-$ and $\alpha-$terms for the same $\beta-$term correspond to a
smaller EMF $\overrightarrow{\varepsilon_{M}}$ and weaker
magnetic flux annihilation. Magnetic flux annihilation, therefore, decreases
from Case E to Case A.
The maximum reconnection electric field $[(\eta+\beta) \vec{J} - \omega
\vec{\Omega} - \alpha \vec{B}]$ in Case E is about 1.3 $E_{0}$ $\sim$ 2.5 $mV/m$.

\begin{figure*}[htbp]
      \centering
          \mbox{
                \includegraphics[width=1.0\textwidth]{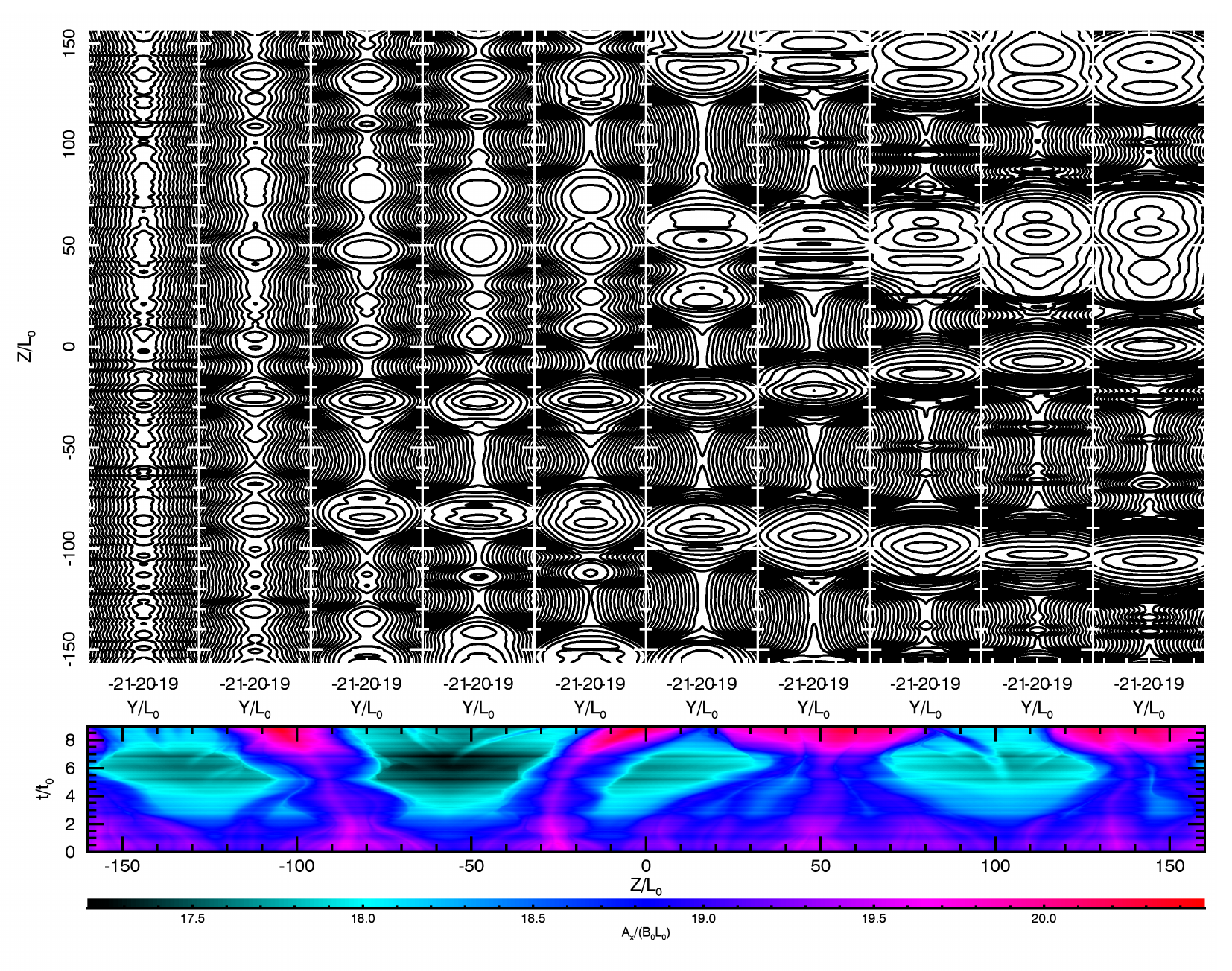}
               }
          \caption{Top: evolution of the mean magnetic field around the midplane
                        ($y=-20 L_{0}$) of the plasmoid-unstable CS.
                  Bottom shows the stack plot of the out-of-plane vector potential component
                        $A_x$ along the midplane of the CS y = -20 $L_{0}$,
                        which is also a proxy of the evolution of the
                  reconnection electric field $E_x$.
                  }
          \label{Fig1}
      \end{figure*}

The results which are presented in the following are obtained for the CS
centered at $y=-d=-20$.
The evolution of the mean magnetic field in this CS is depicted in the top panels of
Fig.~\ref{Fig1}. The first (last) panel depicts the initial (final) CS configuration.
The bottom panel of Fig.~\ref{Fig1} shows a stack plot for the time
evolution of the out-of-plane component of the vector potential $A_{x}$
along the CS center.
Local minima of $A_{x}$ (black and light blue colors) indicate X-points,
while the local maxima of $A_{x}$ (magenta and red colors) correspond to the
centers of plasmoids (O-points).
$A_{x}$ is a proxy of the electric field. Smaller values of $A_{x}$ indicate
stronger electric fields.
With the help of the stack plots in Fig.~\ref{Fig1},
one can identify contraction, coalescence and
expansion of the plasmoids as well as the formation of secondary
plasmoids (indicated by the splitting of black or light-blue regions) in the
course of the nonlinear evolution of the plasmoid
instability.

\section{Electron Acceleration}
\label{Results}

 \subsection{Test particle method}
 \label{Methods}

Due to the overall finite guide magnetic field ($b_g = 2$), a
(relativistic) guiding center approximation can be used to describe the
electron motion.
Its applicability can be proven by calculating the adiabaticity parameter
$\kappa=R_{min}/\rho_{max}$, where $R_{min}$ and $\rho_{max}$ are the minimum
curvature radius of the magnetic fields and the maximum Larmor radius of each
electron \citealp{Buechner1989, Buchner&Zelenyi1991}.
Transition from adiabatic motion to chaotic scattering
is controlled by $\kappa$: for adiabatic electrons, $\kappa$ should be larger than
3. For smaller $\kappa$, the guiding center approximation
is not valid \citealp{Delcourtetal2005}.
In our study, $\kappa$ is always larger than 9 for all electrons,
i.e., the guiding center approximation is appropriate for this investigation.

The relevant guiding center equations of motion to be solved are
\citealp{Northrop1963, Zhouxiaowei2015ApJ, Zhouxiaowei2016ApJ}:
\begin{align}
\displaystyle\frac{d \vec{R}}{dt} = \overrightarrow{v_{D}} &+ \displaystyle\frac{\gamma v_{\parallel}}{\gamma} \vec{b}
\label{GC1} \\
\overrightarrow{v_{D}}  = \overrightarrow{v_{E}}
                           &+ \displaystyle\frac{m}{q} \displaystyle\frac{(\gamma v_{\parallel})^{2}}
                                                                         {\gamma k^{2} B} [\vec{b} \times (\vec{b} \cdot \nabla) \vec{b}]
                            + \displaystyle\frac{m}{q} \displaystyle\frac{\mu }
                                                                         {\gamma k^{2} B} [\vec{b} \times (\nabla(kB))]                        \nonumber\\&
                            + \displaystyle\frac{m}{q} \displaystyle\frac{\gamma v_{\parallel}}
                                                                         {k^{2} B} [\vec{b} \times (\vec{b} \cdot \nabla) \overrightarrow{v_{E}}]
                            + \displaystyle\frac{m}{q} \displaystyle\frac{\gamma v_{\parallel}}
                                                                         {k^{2} B} [\vec{b} \times (\overrightarrow{v_{E}} \cdot \nabla) \vec{b}]        \nonumber\\&
                            + \displaystyle\frac{m}{q} \displaystyle\frac{\gamma}
                                                                         {k^{2} B} [\vec{b} \times (\overrightarrow{v_{E}} \cdot \nabla) \overrightarrow{v_{E}}]
                            + \displaystyle\frac{1}{\gamma c^{2}} \displaystyle\frac{E_{\parallel}}
                                                                         {k^{2} B} (\gamma v_{\parallel}) (\vec{b} \times \overrightarrow{v_{E}})        \nonumber\\&
                            + \displaystyle\frac{m}{q} \displaystyle\frac{\gamma v_{\parallel}}
                                                                         {k^{2} B} (\vec{b} \times \displaystyle\frac{\partial \vec{b}}{\partial t})
                            + \displaystyle\frac{m}{q} \displaystyle\frac{\gamma}
                                                                         {k^{2} B} (\vec{b} \times \displaystyle\frac{\partial \overrightarrow{v_{E}}} {\partial t})
\label{GC2}\\
\displaystyle\frac{d (\gamma v_{\parallel})}{dt} =&   \displaystyle\frac{q}{m} E_{\parallel}
                                - \displaystyle\frac{\mu}{\gamma}[\vec{b} \cdot \nabla B]                                                     \nonumber\\&
                                + (\gamma v_{\parallel}) \overrightarrow{v_{D}} \cdot [ (\vec{b} \cdot \nabla) \vec{b} ]
                                + \gamma \overrightarrow{v_{D}} \cdot [ (\overrightarrow{v_{D}} \cdot \nabla) \vec{b}]                                               \nonumber\\&
                                + \gamma \overrightarrow{v_{D}} \cdot \displaystyle\frac{\partial \vec{b}}{\partial t}
\label{GC3} \\
\gamma =& \sqrt{ \displaystyle\frac{1}{1-(v_{\parallel}^{2}+v_{\perp}^{2}+v_{D}^{2})/c^{2}} }
\label{GC4}\\
\displaystyle\frac{d \mu}{dt} =& 0
\label{GC5}
\end{align}
where $\vec{R}$, $\overrightarrow{v_{D}}$, $v_{\parallel}$, $\gamma$, $\vec{b}$ and $E_{\parallel}$ are the
guiding center position, the perpendicular (to the magnetic field) drift velocity,
the (parallel) velocity along the magnetic field, the relativistic gamma factor
($\gamma= 1 / \sqrt{ 1 -v^{2}/c^{2} }$), the magnetic field unit
vector $\vec{b} = \vec{B} / B $ and the parallel electric field
($\vec{E} \cdot \vec{b} $), respectively.
In Eq.~\eqref{GC2}, $\overrightarrow{v_{E}}$ corresponds to the local $\vec{E} \times \vec{B}$
drift velocity $\overrightarrow{v_{E}} = (\vec{E} \times \vec{B})/B^{2}$.
The other terms of Eq.~\eqref{GC2} describe the magnetic curvature drift velocity,
the gradient drift velocity and so on.
The factor $k = \sqrt{1 - v_{E}^{2} / c^{2} } \simeq 1$ relates the electromagnetic
fields to the reference frame moving with $\overrightarrow{v_{E}}$.
Finally, $\mu = (\gamma v_{\perp})^{2} / (2B) $ is the relativistic
magnetic moment per mass unit with $v_{\perp}$ being the particle's gyration
velocity in the direction perpendicular to $\vec{B}$.
The electron kinetic energy can be expressed using the relativistic $\gamma$-factor as
$E_{k}=(\gamma-1)mc^{2}$.
We solve Eqs.~\eqref{GC1} to \eqref{GC5} utilizing a
fourth-order Runge-Kutta scheme.
The electromagnetic fields obtained by the MHD simulations
are linearly interpolated to the electron positions between the grid points.

Due to the small non-relativistic drift speeds
$\overrightarrow{v_{D}} \approx \overrightarrow{v_{E}}
< 2 V_{0}$, the electron kinetic energy $E_{k}$ is proportional to
$v_{\parallel}^{2}+ v_{\perp}^{2}+v_{D}^{2} \simeq v_{\parallel}^{2}+ v_{\perp}^{2}$.
The resulting energy change rates in the parallel and perpendicular
directions are given, respectively, by:
\begin{align}
      \displaystyle\frac{1}{2}\displaystyle\frac{d (\gamma v_{\parallel})^{2}}{dt} &=
                                 \displaystyle\frac{q}{m} \gamma v_{\parallel} E_{\parallel}
                                - \mu v_{\parallel}(\vec{b} \cdot \nabla B)                                               \nonumber\\&
                                + (\gamma v_{\parallel})^{2} \overrightarrow{v_{D}} \cdot [ (\vec{b} \cdot \nabla) \vec{b} ]
                                + \gamma^{2} v_{\parallel} \overrightarrow{v_{D}} \cdot [ (\overrightarrow{v_{D}} \cdot \nabla) \vec{b}]  \nonumber \\&
                                + \gamma^{2} v_{\parallel} \overrightarrow{v_{D}} \cdot \displaystyle\frac{\partial \vec{b}}{\partial t}
\label{Parallel_Gain}\\
      \displaystyle\frac{1}{2}\displaystyle\frac{d (\gamma v_{\perp})^{2}}{dt} &= \displaystyle\frac{d \mu B}{dt} = \mu \displaystyle\frac{d B}{dt}   \nonumber\\&
                               = \mu \displaystyle\frac{\partial B}{\partial t}
                               + \mu v_{\parallel}(\vec{b} \cdot \nabla B)
                               + \mu \overrightarrow{v_{D}} \cdot \nabla B
\label{Perpendicular_Gain}
\end{align}
And the total rate of kinetic energy change is given by:
\begin{align}
      \displaystyle\frac{d E_{k} }{dt} &= \vec{v} \cdot \displaystyle\frac{d \vec{P}}{dt}
                                        = \displaystyle\frac{m}{2 \gamma}\displaystyle\frac{ d(\gamma v)^2 }{dt}                                     \nonumber\\
                                       &\cong \displaystyle\frac{m}{\gamma} \displaystyle\frac{1}{2}\displaystyle\frac{d (\gamma v_{\parallel})^{2} +(\gamma v_{\perp})^{2}}{dt} \nonumber\\
                                       &=  q v_{\parallel} E_{\parallel}
                                        + (m \gamma v_{\parallel}^{2}) \overrightarrow{v_{D}} \cdot [ (\vec{b} \cdot \nabla) \vec{b} ]                \nonumber\\&
                                        + (m \gamma v_{\parallel}) \overrightarrow{v_{D}} \cdot [ (\overrightarrow{v_{D}} \cdot \nabla) \vec{b}]
                                        + \displaystyle\frac{m \mu}{\gamma} \overrightarrow{v_{D}} \cdot \nabla B                                      \nonumber\\&
                                        + (m \gamma v_{\parallel}) \overrightarrow{v_{D}} \cdot \displaystyle\frac{\partial \vec{b}}{\partial t}
                                        + \displaystyle\frac{m \mu}{\gamma} \displaystyle\frac{\partial B}{\partial t}
\label{Total_Gain}
\end{align}
As one can see in Eq.~\eqref{Total_Gain}, electrons can gain energy
by parallel electric fields ($q v_{\parallel} E_{\parallel}$ term),
due to magnetic field curvature ($m \gamma v_{\parallel}^{2} \overrightarrow{v_{D}} \cdot [ (\vec{b} \cdot \nabla) \vec{b} ]$ term),
by curvature drift acceleration ($m \gamma v_{\parallel} \overrightarrow{v_{D}} \cdot [ (\overrightarrow{v_{D}} \cdot \nabla) \vec{b}]$ term),
by (perpendicular) gradient acceleration ($m \mu (\overrightarrow{v_{D}} \cdot \nabla B)/\gamma$ term),
and temporal variation of the magnetic field, here
$m \gamma v_{\parallel} \overrightarrow{v_{D}} \cdot (\partial \vec{b}/\partial t)$ and $m \mu (\partial B /\partial
t)/\gamma$ terms for the parallel and perpendicular acceleration, respectively.
The parallel magnetic field gradient $(m \mu) [v_{\parallel}(\vec{b} \cdot \nabla
B)]/\gamma$ can change the parallel and perpendicular energies simultaneously
and transfer energy between them while it does not change the total particle
energy.

For our calculations, we inject $1.7 \times 10^{5}$ electrons at $t=0$
randomly distributed along the midplane of the CS (see the first panel of Fig.~\ref{Fig1}).
The cosines of electrons' initial pitch angle randomly vary in the range of $[-1, 1]$.
The initial energies are uniformly distributed between 10\,eV and 10\,keV.
Electrons are traced until 9 $t_{0}$ ($\sim 9 s$), the typical
duration of electron bursts in the Hermean magnetotail.

\subsection{Acceleration features}
\label{Acceleration_features}

\begin{figure*}[htbp]
      \centering
          \mbox{
                \includegraphics[width=1.0\textwidth]{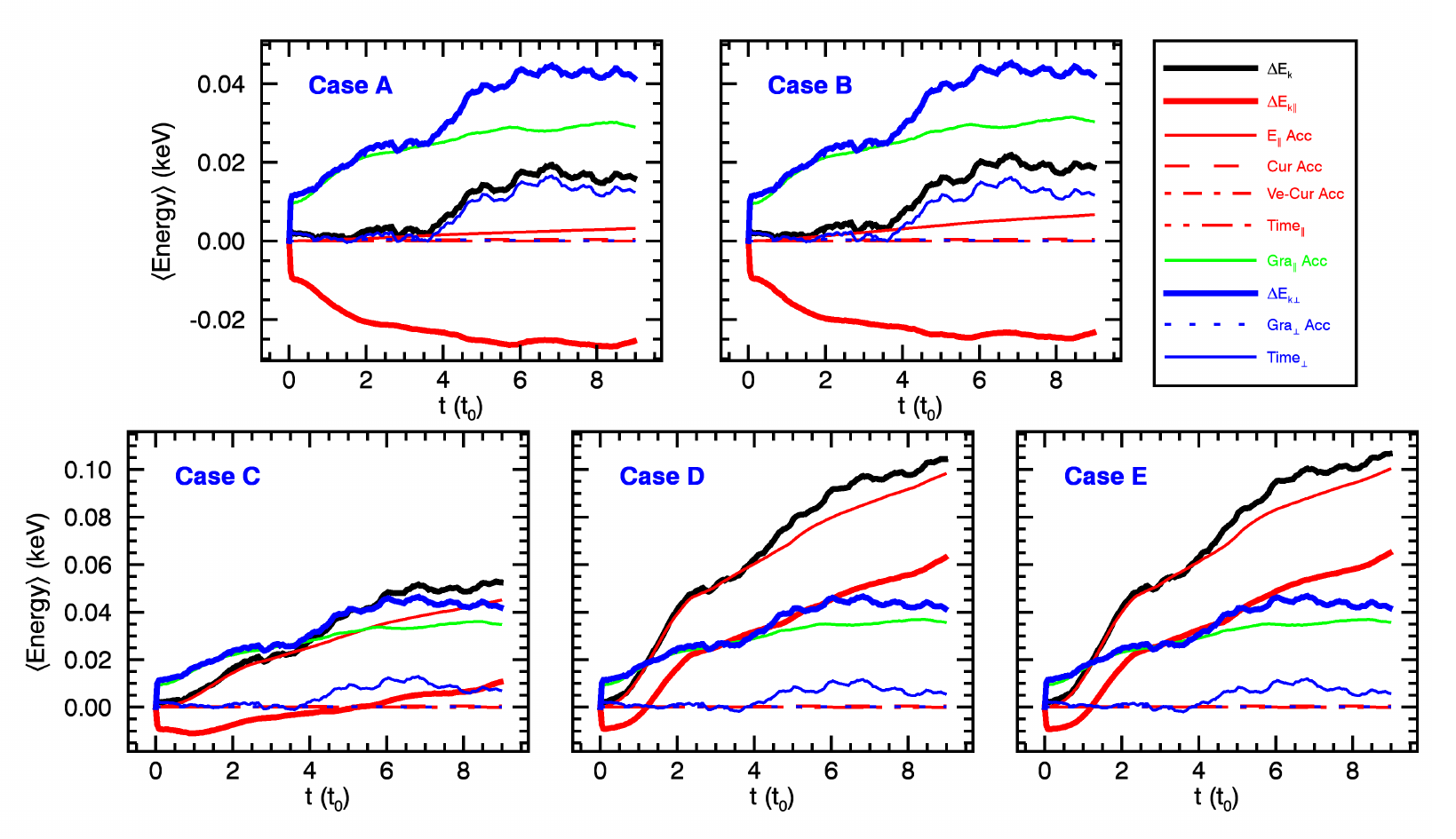}
               }
          \caption{Average contribution of the different acceleration
                   terms to the electron energy gain:
                   Total energy $\Delta E_{k}=E_{kt}-E_{k0}$ (thick black lines),
                   gain in the parallel direction $\Delta E_{k \parallel}=E_{k \parallel t}-E_{k \parallel 0}$ (thick red lines),
                   perpendicular energy gain $\Delta E_{k\perp}=E_{k \perp t}-E_{k \perp 0}$ (thick blue lines).
                   The acceleration terms are:
                   parallel electric field acceleration (thin red lines),
                   magnetic curvature acceleration (red dashed lines),
                   curvature drift acceleration (red dash-dot lines),
                   time change effects in the parallel direction (red dash-dot-dot-dot lines),
                   parallel gradient acceleration (cyan lines),
                   perpendicular gradient acceleration (blue dotted lines)
                   as well as magnetic field variation acting in the perpendicular direction (thin blue lines).
                   Note the different range in the y-axis between the panels of the first and second rows.
                   }
          \label{Acceleration_Evolution}
      \end{figure*}

Fig.~\ref{Acceleration_Evolution} depicts the temporal evolution
of the different contributing acceleration mechanisms of electrons in
dependence on the turbulence parameters (see Table~\ref{tab1}).
In the laminar reconnection limit (Case A), resistive
electric fields $E_{\parallel}=\eta J_{{\parallel}}$ do not significantly
accelerate electrons.
Instead, mainly the temporal variation of the magnetic field
(the $\mu(\partial B /\partial t)/\gamma$ term) energizes
electrons.
In case of weak turbulence (small coefficient $C_{\beta}$, Case B),
the $E_{\parallel}$ acceleration is slightly enhanced but it is still smaller
than the acceleration by $\partial B /\partial t$.
The overall acceleration in Case B is as efficient as in Case A.
The parallel kinetic energy $E_{k \parallel}$ is reduced in both Cases A and B
due to the effect of the parallel gradient
$\mu [v_{\parallel}(\vec{b} \cdot \nabla B)]/\gamma$, which
transfers kinetic energy from the parallel to the perpendicular electron
motion.

Stronger turbulence (larger coefficients
$C_{\beta}$ and/or $\tau$, Cases C, D and E) acts as enhanced
localized anomalous resistivity as well as electron acceleration via the
larger EMF $\overrightarrow{\varepsilon_{M}}$, where electron acceleration
is mainly due to the parallel electric field ($E_{\parallel}$).
By comparing Cases C and E, one can find that electron acceleration is enhanced
with the enhancement of the turbulence ($C_{\beta}$).
The difference between Cases D and E is that the $\omega-$ and $\alpha-$terms
(see Eq.~\eqref{ElectricField}) in Case D are two times as large as
in Case E, while the $\beta-$term contribution remains unchanged.
Larger $\omega-$ and $\alpha-$terms indicate weaker magnetic flux
annihilation, and therefore the effects of the $\beta$-term and EMF
$\overrightarrow{\varepsilon_{M}}$ in Eq.~\eqref{ElectricField}
are reduced (Sect.~\ref{MHD-Model}).
Comparing with Case D, hence, electrons in Case E are accelerated to
higher energies $\sim$ 70 keV.

In any case, in addition to the acceleration by the parallel electric fields
($E_{\parallel}$), acceleration effects due to $\partial B /\partial t$ and
$v_{\parallel}(\vec{b} \cdot \nabla B)$ are important, which lead to electron
energization in the direction perpendicular to the magnetic field
(see Eq.~\eqref{Perpendicular_Gain}).
Comparing with the $E_{\parallel}$ acceleration, $\partial B /\partial t$ and
$v_{\parallel}(\vec{b} \cdot \nabla B)$ accelerations depend, however, very weakly
on the turbulence level.
Due to the enhanced parallel velocity $v_{\parallel}$ by $E_{\parallel}$ acceleration,
the energy gain from the $\partial B /\partial t$ ($v_{\parallel}(\vec{b} \cdot \nabla B)$)
term slightly decreases (increases) from Case A to E.
The curvature
$\overrightarrow{v_{D}} \cdot [ (\vec{b} \cdot \nabla) \vec{b}]$,
perpendicular gradient $(\overrightarrow{v_{D}} \cdot \nabla B)/\gamma$,
curvature drift
$\overrightarrow{v_{D}} \cdot [ (\overrightarrow{v_{D}} \cdot \nabla) \vec{b}]$
and perpendicular gradient $(\overrightarrow{v_{D}} \cdot \nabla B)/\gamma$
accelerations practically do not contribute to the electron energization.

As a result, for negligible or weak turbulence levels,
electrons are mainly accelerated by the temporal variation of the magnetic field
($\partial B /\partial t$ term) in the chain of plasmoids.
In cases of stronger turbulence, however, the parallel
electric field ($E_{\parallel}$) acceleration due to the localized
EMF $\overrightarrow{\varepsilon_{M}}$ dominates the electron
energization.

 \subsection{Acceleration sites}   

\begin{figure*}[htbp]
      \centering
          \mbox{
                \includegraphics[width=1.0\textwidth]{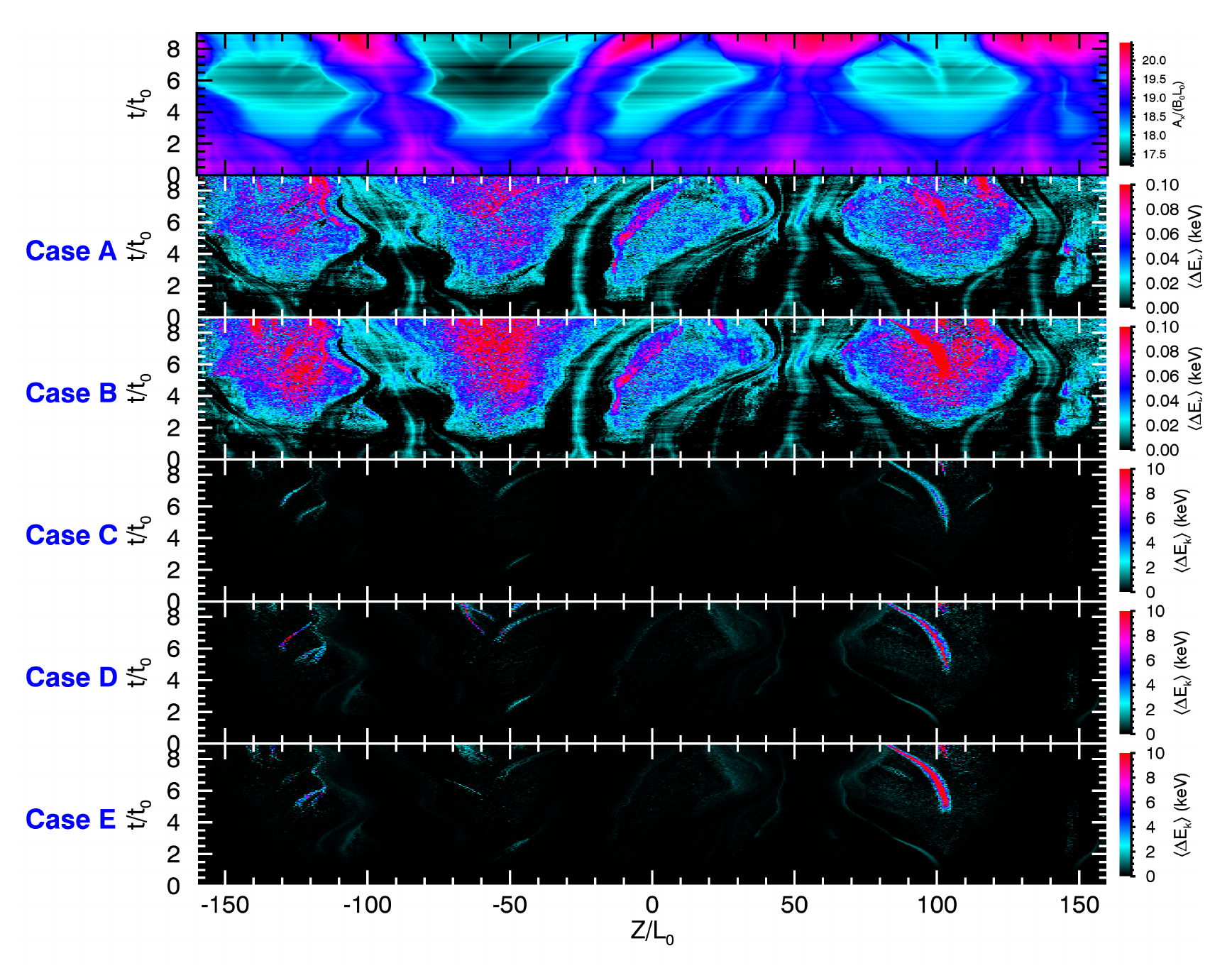}
               }
          \caption{
                   Top: $A_{x}$ stack plot as at the bottom of Fig.~\ref{Fig1}, indicating the location of
                        X- and O-lines inside plasmoids.
                   Bottom five panels: electron total kinetic energy gain ($\Delta E_{k} =
                   E_{kt}-E_{k0}$) versus the Z-axis projection of the electron positions at the time $t$,
                   in order to visualize their acceleration sites with respect to the X-points and plasmoids.
                   Note the different range in the color scheme between Cases A, B compared with C, D and E.
                  }
          \label{Z_EnergyGain_Evolution_Simple}
      \end{figure*}

The bottom five panels of Fig.~\ref{Z_EnergyGain_Evolution_Simple} display
the time evolution of the electron total kinetic energy gain
($\Delta E_{k} = E_{kt}-E_{k0}$) versus the (Z-axis-) position of the electrons
for the five different turbulence levels (Cases A to E).
The plots localize the electron acceleration sites with respect to
the X-points and plasmoid centers (see the stack plot of the vector
potential $A_{x}$ in the top panel).
As one can see the electrons are energized mainly around X-points
(black and light blue regions in the top panel), even though the dominant
electron acceleration mechanisms are different depending on the turbulence level.
That is due to the injection of new reconnecting magnetic flux and strong
current density near the X-points to enhance the
$\partial B /\partial t$ and $E_{\parallel}$ acceleration, respectively.
Only a small amount of energy is gained inside the plasmoids, this
energization is also mainly due to the temporal variation of
the magnetic field $\partial B /\partial t$. Acceleration by
$\partial B /\partial t$ in the plasmoids is, however, much weaker than the
acceleration processes taking place near the X-points.

The bottom five panels of Fig.~\ref{Z_EnergyGain_Evolution_Simple} illustrate
the formations of filamentary structures in the spatial distribution of the
energized electrons.
The number of the filaments increases with the enhancement of the turbulence.
The filaments start at X-points firstly and move then into a channel
formed by the split of an X-point (see the top panel).
Splitting of X-points indicates the formation of new plasmoids.
Therefore, along the
filamentary structures, electrons are first getting accelerated at an X-point
and then trapped inside the newly formed and moving secondary plasmoids
\citealp{Bartaetal2011}.

\begin{figure*}[htbp]
      \centering
          \mbox{
                \includegraphics[width=0.8\textwidth]{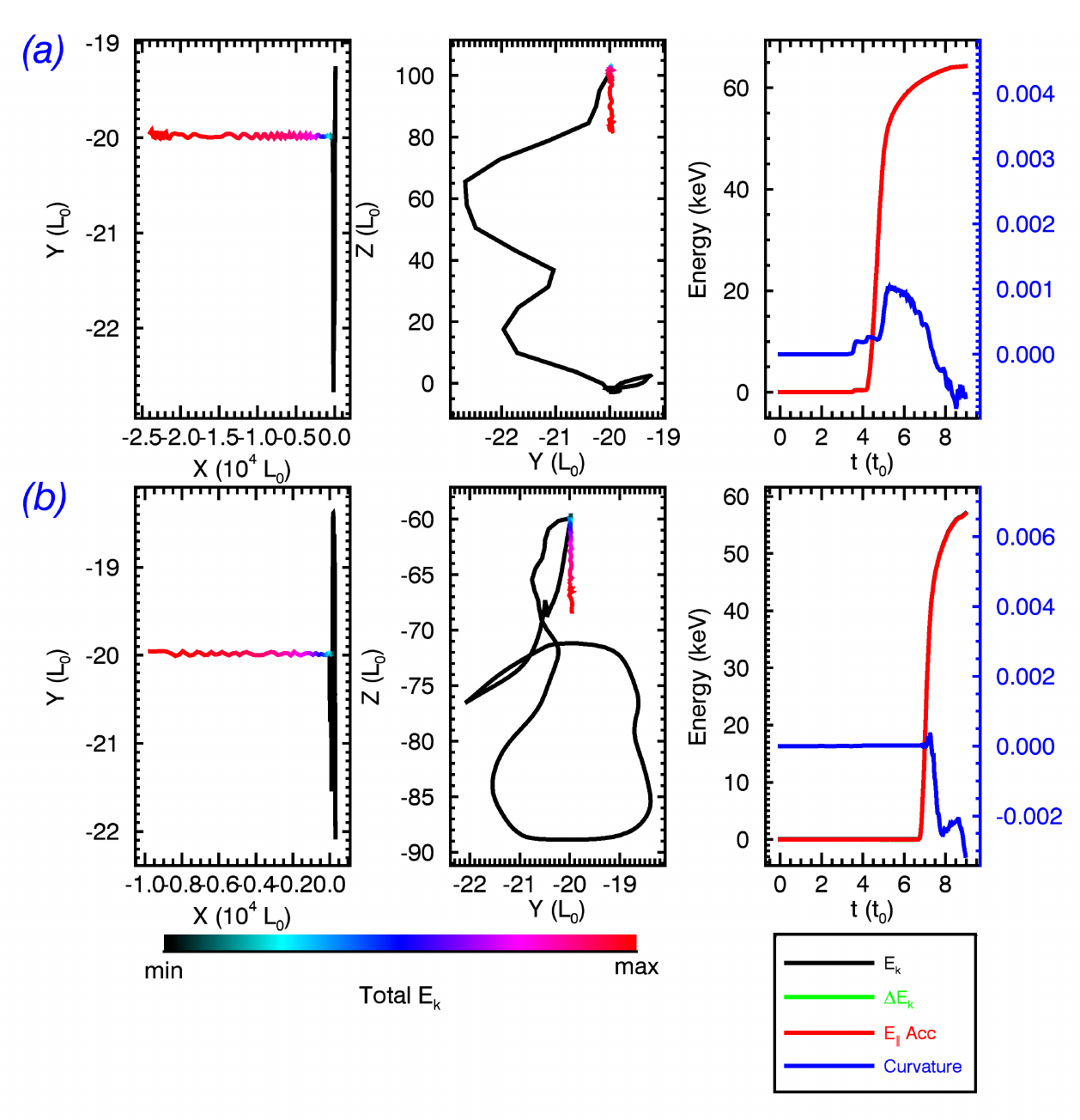}
               }
          \caption{Trajectory and energy evolution of two characteristic electrons "(a) and (b)" in strong turbulence (Case D).
                   First and second column: projection of the electron trajectories in the $xy$ and $yz$ planes. These
                   trajectories are color-coded by their total kinetic energy $E_{k}$ profiles.
                   Third column: evolution of $E_{k}$ (black line), total energy gain $\Delta
                   E_{k}$ (cyan line), parallel electric field acceleration ($E_{\parallel}$ Acc, red
                   line) and acceleration due to the curvature acceleration (Curvature, blue
                   line). The y-axis in the right side corresponds only to the curvature acceleration.
                   In the animations, the first column shows the location evolution of the electron (red cross)
                   in the CS, the second colum is similar to the panels in the third column of
                   Fig.\ref{Trajectory}. Vertical black dashed line in the second colum of the animations corresponds to
                   the time of the first column.
                  }
          \label{Trajectory}
      \end{figure*}

Fig.~\ref{Trajectory} illustrates this process by showing the
trajectory and energy evolution of two characteristic electrons
in Case D (similar in Case E).
The two electrons are first accelerated by the parallel electric
fields ($E_{\parallel}$) at the X-points near $Z=100$ and -60 $L_{0}$
respectively (see the top and bottom panels).
Their accelerations take place mainly at the center of the CS ($Y=-20$ $L_{0}$).
Since these two electrons are accelerated by $E_{\parallel}$ at different
times $t$, the filamentary structures near $Z=100$ and -60 $L_{0}$ appear at
different moments of time (see Fig.~\ref{Z_EnergyGain_Evolution_Simple}).
Meanwhile the electrons near $Z=100$ $L_{0}$ spend a longer time in
the parallel electric field ($E_{\parallel}$) than those accelerated
near $Z=-60$ $L_{0}$.
Therefore, the overall energy gain of the electrons accelerated near
$Z=-60$ $L_{0}$ is smaller than that of the electrons accelerated near
$Z=100$ $L_{0}$ despite of the fact that the parallel electric fields
($E_{\parallel}$) are stronger near $Z=-60$ $L_{0}$ (see the slopes
of the energy evolution in the third column of Fig.~\ref{Trajectory}).

Although the parallel electric field ($E_{\parallel}$) acceleration dominates the
electron energization, it is also interesting to pay attention to their curvature
($\overrightarrow{v_{D}} \cdot [ (\vec{b} \cdot \nabla) \vec{b} ]$ term)
energization, which leads to the small deceleration of electrons in this study.
Animations of their trajectories clearly
show that the electrons are affected by the expansion and coalescence
of the secondary plasmoids in which they are trapped.
At the beginning, the parallel electric field energization enhances the
curvature acceleration, leading to its maximum
(blue lines in the third column of Fig.~\ref{Trajectory}) near an X-point.
Later, however, the electrons are removed from this region
along the CS by the newly formed expanded secondary plasmoids.
The plasmoid expansion causes
a deceleration of the electrons via the curvature term.
Almost at the end (after 8 $t_{0}$),
plasmoid coalescence again causes electron acceleration which is, however,
much smaller than the energy lost due to the plasmoid expansion.
Hence, instead of accelerating, the curvature
$\overrightarrow{v_{D}} \cdot [ (\vec{b} \cdot \nabla) \vec{b}]$ term
decelerates these electrons by four orders of
magnitude less than the $E_{\parallel}$ acceleration.

Previous studies concluded that the
curvature $\overrightarrow{v_{D}} \cdot [ (\vec{b} \cdot \nabla) \vec{b}
]$ in the plasmoids can significantly contribute to the particle
energization at large (MHD) scales \citealp{Zhouxiaowei2016ApJ}.
We found, however, that the contribution of the curvature
$\overrightarrow{v_{D}} \cdot [ (\vec{b} \cdot \nabla) \vec{b}]$ term is negligibly small.
As Fig.~\ref{Fig1} demonstrates the plasmoid expansion
dominates the reconnection process despite of multiple contractions and
coalescences of the plasmoids.
That is the reason for the weak contribution of the curvature
$\overrightarrow{v_{D}} \cdot [ (\vec{b} \cdot \nabla) \vec{b} ]$
acceleration in turbulent plasmoid reconnection (Fig.~\ref{Acceleration_Evolution}).

\subsection{Energy spectra}

\begin{figure*}[htbp]
      \centering
          \mbox{
                \includegraphics[width=1.0\textwidth]{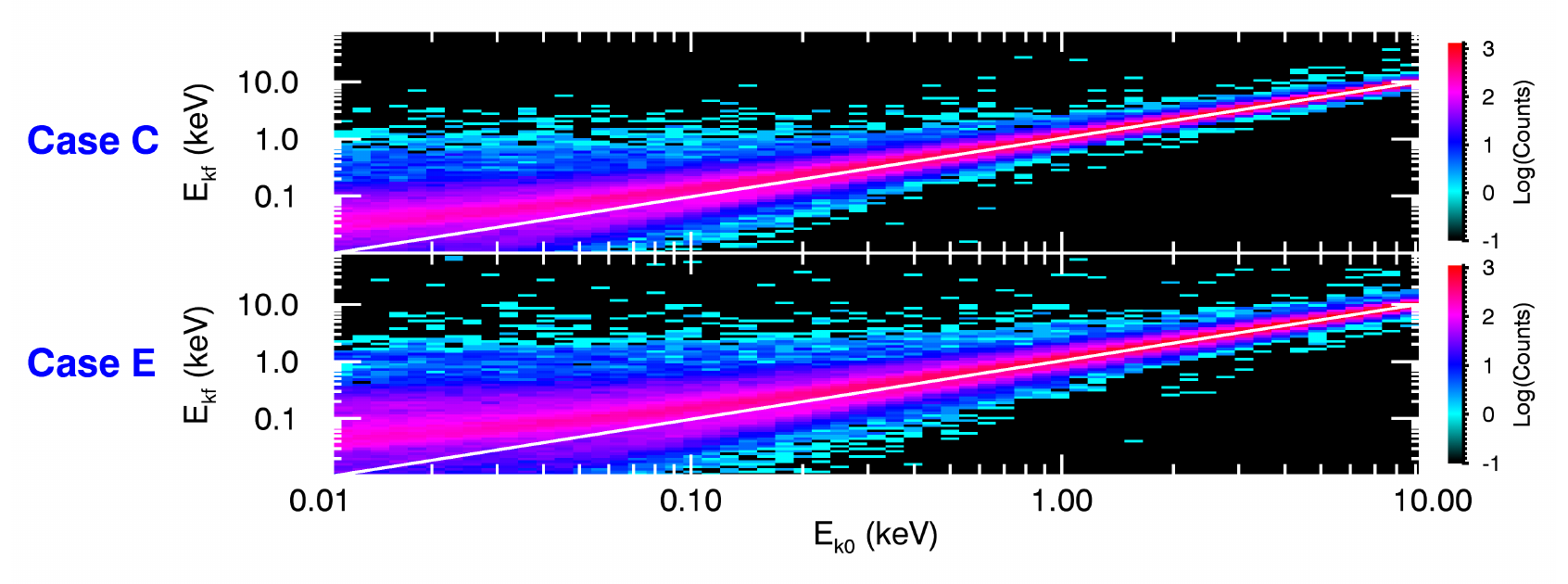}
               }
          \caption{Number distribution of energetic electrons in the initial-final
                   kinetic energy ($E_{k0}-E_{kf}$) space for the strong turbulence Case C and E.
                   White lines correspond to "$E_{kf}=E_{k0}$", separating regions of
                   electron net acceleration and deceleration.
                  }
          \label{Count_Distribution}
      \end{figure*}

Fig.~\ref{Count_Distribution} depicts the number distribution of energetic
electrons in the space of the initial ($E_{k0}$) and final ($E_{kf}$) kinetic
energy in cases of strong turbulence (Cases C and E).
White lines $E_{k0}=E_{kf}$ divide accelerated (above $E_{kf}=E_{k0}$) and
decelerated (below $E_{kf}=E_{k0}$) electrons.
Due to the small EMF $\overrightarrow{\varepsilon_{M}}$ in Cases A and B, most
electrons in these two cases maintain their initial energies and distribute
near the line $E_{kf}=E_{k0}$. The electron distribution in Cases D and
E, on the other hand, are similar, so Case D is not shown in
Fig.~\ref{Count_Distribution}. Kinetic energy $E_{kf}$ of most
electrons in Cases C, D and E stays below 20 keV, a smaller number of electrons can be
accelerated up to 60 keV.
In these three cases, the most energetic electrons are homogeneously distributed
in the $E_{k0}$ space. This means that the electrons are accelerated to higher
energies independent on their initial energies $E_{k0}$ \citealp{Zhouxiaowei2016ApJ}.

\begin{figure*}[htbp]
      \centering
          \mbox{
                \includegraphics[width=1.0\textwidth]{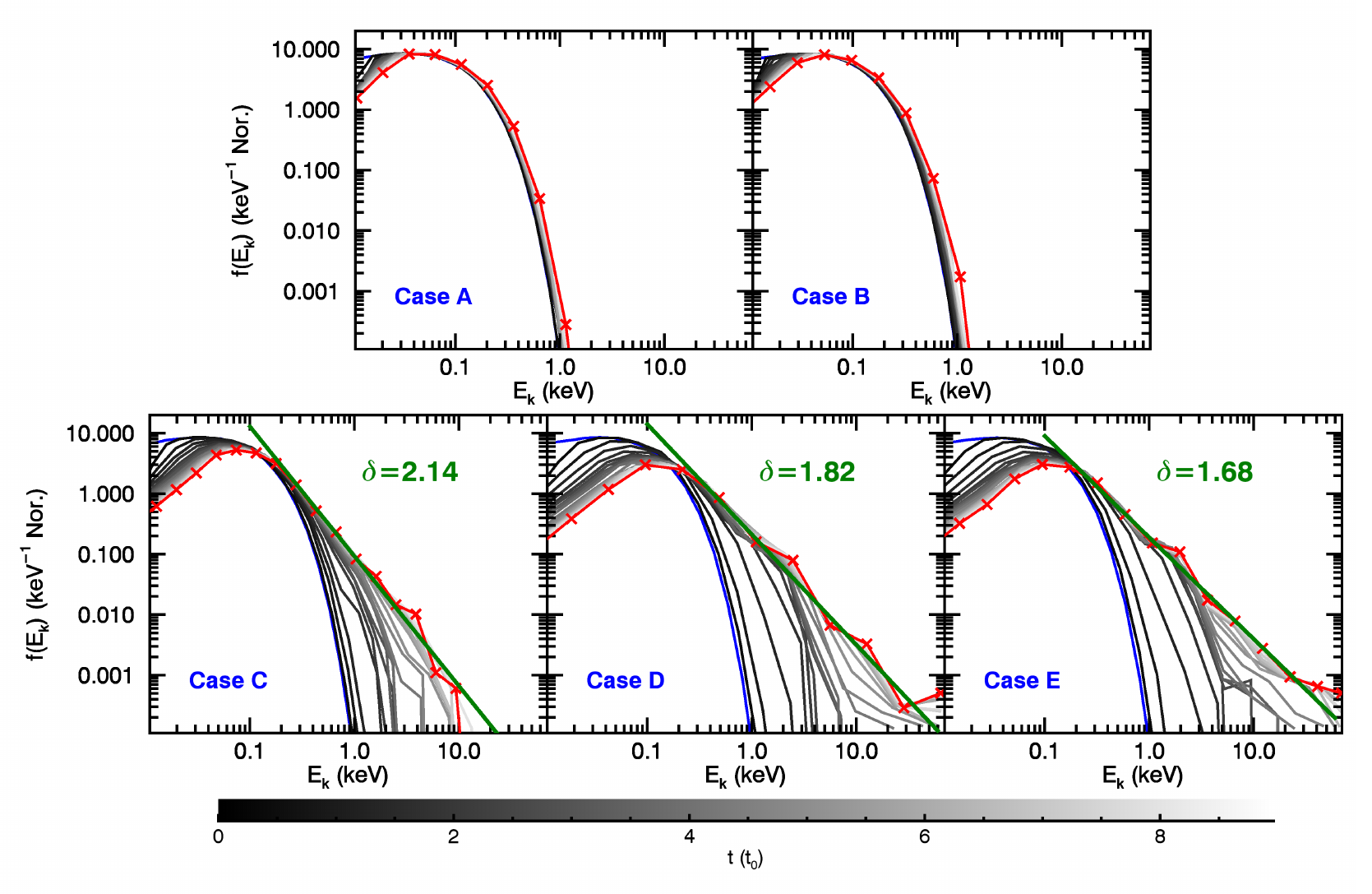}
               }
          \caption{Electron spectra dependence on the strength of the turbulence. Grayscale is
                   used to distinguish spectra at different time $t$ (from black to bright). Initial
                   and final electron spectra are highlighted by
                   blue and red lines, respectively. Green lines in the
                   bottom panels are power law fittings with spectral index $\delta$ given in the
                   corresponding panel.
                  }
          \label{Spectrum}
      \end{figure*}

The five panels of Fig.~\ref{Spectrum} show the evolution of the electron
energy spectra for the five different turbulence levels.
We calculate the energy spectra by assigning a statistical weight $\psi(E_{k})$
to each electron. The weight function $\psi(E_{k})$ depends on the
initial electron energy $E_{k0}$ as
$\psi(E_{k})=\sqrt{E_{k0}} \exp(- E_{k0} / E_{th})$. We choose
$E_{th} \sim 0.06 keV$ to match with the typical electron thermal energy
observed in the Hermean magnetotail. The initial energy distribution is, therefore,
Maxwellian (shown as a blue line in each panel of
Fig.~\ref{Spectrum}).
One can see that the final electron energy spectra strongly
depend on the strength of turbulence.
In accordance with the results discussed in Sect.~\ref{Acceleration_features},
the spectra of electrons energized in weakly turbulent CSs
(negligible or small EMF $\overrightarrow{\varepsilon_{M}}$,
Cases A and B) do not differ much from the initial Maxwellian distribution.
In these cases, electrons are only slightly heated due to the (small)
background resistivity $\eta$.
In contrast, for strong turbulence (larger EMF $\overrightarrow{\varepsilon_{M}}$,
Cases C, D and E),
power-law spectra are formed at energies above 100 eV with a spectral
index $\delta \sim$  2.1 in Case C and 1.7 - 1.8 in Cases D and
E.
The stronger turbulent EMF $\overrightarrow{\varepsilon_{M}}$, the harder
the spectrum of energetic electrons becomes.

\section{Conclusions and discussions}
\label{Conclusions}

Recently, \emph{Widmer et al.} \citealp{Widmeretal2016PhPl} revealed how
turbulence can enhance reconnection rates via the electromotive force
(EMF $\overrightarrow{\varepsilon_{M}}$) with the mean-field turbulence model
of \emph{Yoshizawa} \cite{Yoshizawa1990PhFlB} and
\emph{Yokoi $\&$ Hoshino} \cite{Yokoi&Hoshino2011PhPl}.
We now found that in strong turbulence, the turbulence driven
EMF $\overrightarrow{\varepsilon_{M}}$
can also cause efficient electron acceleration via the enhanced localized
parallel electric fields ($E_{\parallel}$) near X-points here (Cases C, D and E).
Power-law spectra up to 60 keV form out of an initial Maxwellian distribution with a
thermal electron energy 60 eV typical for the Hermean magnetotail.
If other typical parameters of the Hermean magnetotail are considered, the
spectral indices of the power-law spectra are between 1.68 - 2.14.
The higher the turbulence level, the harder the electron spectrum.
Indeed, energetic electron distributions in the Hermean magnetotail
observed by MESSENGER often show power-law indices between 1.5 and 4
\citealp{Ho2011a, Hoetal2012JGRA, Lawrenceetal2015JGRA}.
Turbulence contributes to the
electron energization via the EMF $\overrightarrow{\varepsilon_{M}}$, in
a similar way as a localized anomalous resistivity
\citealp{Gordovskyyetal2010a, Yang2015RAA}.
On the other hand, for weak turbulence (negligible or small EMF
$\overrightarrow{\varepsilon_{M}}$, here Cases A and B), the rate of
the change of the magnetic-field ($\partial B /\partial t$) dominates the
electron energization, but the electrons just become heated and not
significantly accelerated.

Previous studies
\citealp{Turkmanietal2006, Gordovskyyetal2010b, Zhouxiaowei2016ApJ} have
found that electron acceleration is dominated by parallel electric
field ($E_{\parallel}$) assuming large "anomalous" resistivities.
We found that strong acceleration is mainly due to a localized turbulent EMF
$\overrightarrow{\varepsilon_{M}}$.
Parallel electric fields ($E_{\parallel}$) in strong turbulence
and magnetic field variations ($\partial B /\partial t$) both accelerate
electrons mainly near the reconnection X-lines, where current density and
compression of the magnetic flux reach their maxima.
$\partial B /\partial t$ term can accelerate electrons also inside plasmoids,
but its overall efficiency is much
weaker than the acceleration near X-points.
Curvature $\overrightarrow{v_{D}} \cdot [ (\vec{b} \cdot \nabla) \vec{b} ]$
acceleration due to the contraction and coalescence of
plasmoids, however, does  not contribute significantly to
the electron acceleration in case of turbulent plasmoid reconnection, since
the plasmoid expansion dominates the plasmoid reconnection process, slowing down
the electrons.

Our results provide evidence that turbulence not
only enhances the rate of magnetic reconnection but also efficiently accelerates
electrons if the turbulence is strong enough.
This complements, e.g., the study of \emph{Kowal et al.} \cite{Kowaletal2012PhRvL},
who used a different turbulence model applied to the acceleration processes
in the interstellar medium (ISM).
Our findings also complement previous studies
\citealp{Delcourtetal2005, Schriver2011, Walsh2013}
of particle acceleration in the Hermean magnetotail, which did not
consider turbulence, while turbulence is ubiquitous in collisionless plasmas.

Note that the parametrizations of the turbulence by the
coefficients $C_{\beta}$ and $\tau$ is  based on previous studies.
In order to address the effects of the turbulence on electron acceleration in
plasmoid-unstable CSs and test the electron acceleration results in this paper,
more detailed self-consistent studies with the mean-field
turbulence model will have to be carried out in the future.
Meanwhile the parametrizations of the turbulence for which we found the most
efficient electron acceleration may serve as a reference for future
detailed studies of the turbulent plasmoid reconnection.
At least they provide upper limits for electron acceleration by
the turbulent plasmoid reconnection which can be expected to be measured by
the Bepi Colombo mission to Mercury to be launched soon.

%
\begin{acknowledgements}
X. Zhou thanks the Max-Planck-Society for  granting her a postdoc-stipend.
F. Widmer acknowledges the IMPRS School at the Max Planck Institute for Solar
System Research for the financial support at the time of this work.
J. B\"{u}chner and P. A. Mu\~{n}oz acknowledge financial support from the
Max-Planck-Princeton Center for Plasma Physics
and the DFG Schwerpunktprogramm "PlanetMag" SPP 1488.
\end{acknowledgements}

\pagebreak
\clearpage
%
%

%
%
\pagebreak
\clearpage
%
%
%
%
\pagebreak
\clearpage
%
%
%
%
\pagebreak
\clearpage
%
%
%
%
\pagebreak
\clearpage
%
%
%
%
\pagebreak
\clearpage
%
%

\end{document}